\documentclass[%
 reprint,
superscriptaddress,
 amsmath,amssymb,bm
prapplied
]{revtex4-2}

\usepackage{graphicx}
\usepackage{dcolumn}
\usepackage{bm}
\usepackage{booktabs}
\usepackage{amsmath}
\begin{document}


\title{Electromagnetic Nonreciprocity in a Magnetized Plasma Circulator}

\author{Feng Li}
\email{feli@ucsd.edu}
\affiliation{Electrical and Computer Engineering Department, University of California, San Diego, La Jolla, California 92093, USA}

\author{Robert J. Davis}
\email{rjdavis@ucsd.edu}
\affiliation{Electrical and Computer Engineering Department, University of California, San Diego, La Jolla, California 92093, USA}

\author{Sara M. Kandil}
\email{skandil@ucsd.edu}
\affiliation{Electrical and Computer Engineering Department, University of California, San Diego, La Jolla, California 92093, USA}

\author{Daniel F. Sievenpiper}
\email{dsievenpiper@ucsd.edu}
\affiliation{Electrical and Computer Engineering Department, University of California, San Diego, La Jolla, California 92093, USA}

\date{\today}

\begin{abstract}
Nonreciprocal transport of electromagnetic waves within magnetized plasma is a powerful building block towards understanding and exploiting the properties of more general topological systems. Much recent attention has been paid to the theoretical issues of wave interaction within such a medium, but there is a lack of experimental verification that such systems can be viable in a lab or industrial setting. This work provides an experimental proof-of-concept by demonstrating nonreciprocity in a unit component, a microwave plasma circulator. We design an E-plane Y junction plasma circulator operating in the range of 4 to 6 GHz using standardized waveguide specifications. From both simulations and experiments, we observe wide band isolation for the power transmission through the circulator. The performance and the frequency band of the circulator can be easily tuned by changing the plasma density and the magnetic field strength. By linking simulations and experimental results, we estimate the plasma density for the device.

\end{abstract}

\maketitle

\section{Introduction}

The interaction between electromagnetic waves and magnetized plasmas has been studied for decades \cite{ginzburg_textbook_1970}, but only recently has it been recognized that they share a deep topological connection \cite{silveirinha_chern_2015}. Topological electromagnetics generally hinges upon discretely translatable systems, where the closed nature of the Brillouin zone permits the definition of the Chern number, which in turn provides a nonreciprocal band structure protected by the global topology \cite{Bisharat_2021}. The strongest form of topological protections (e.g., systems emulating the integer quantum Hall effect) requires nonreciprocity, which can prove challenging to realize in electromagnetic systems. The use of ferrite materials has become standard for use in nonreciprocal devices, owing to the off-diagonal elements of the material's permeability tensor influencing transverse-magnetic (TM) modes \cite{haldane_possible_2008,raghu_analogs_2008,wang_reflection-free_2008,wang_observation_2009}. For a plasma with an applied perpendicular magnetic field, off-diagonal terms also appear within the material's permittivity tensor (the Voigt effect), influencing transverse-electric (TE) modes \cite{boyd_textbook_2007}. This has led to the proposal of a new class of nonreciprocal systems based on plasmas, specifically those in the experimentally relevant area of photonic crystals \cite{hojo_dispersion_2004}. Prior studies have numerically investigated the band diagram of 1D to 3D plasma photonic crystals with and without magnetization using methods including finite-difference-time-domain (FDTD), transfer matrix method, or plane wave expansion method \cite{kuzmiak_photonic_1994,zhang_study_2009,qi_dispersion_2010,zhang_photonic_2012,qi_properties_2010,qi_photonic_2012,fathollahi_khalkhali_effect_2016}. A number of experimental works have verified wave interaction with periodic arrays of unmagnetized plasma rods \cite{sakai_verification_2005,sakai_interaction_2005,sakai_characteristics_2007} and demonstrated tunable plasma photonic crystal filters \cite{wang_tunable_2015, wang_plasma_2016}.

Nonreciprocity has long existed in non-periodic ferrite systems, such as circulators, isolators, and other energy-directing devices. It has recently been shown that bulk plasma likewise possesses nontrivial band topology \cite{silveirinha_chern_2015, hassani_gangaraj_berry_2017,parker_topological_2020, parker_topological_2021,fu_topological_2021}, even in the absence of periodicity. This interaction between nonreciprocity and topology can thus be readily explored within magnetized plasma systems, where it is possible to engineer both a bulk nonreciprocal response as well as a non-trivial band structure. Previous works have numerically studied nonreciprocal propagation and wave-routing behavior via bulk plasma \cite{yang_one-way_2016,xi_polarization-independent_2018}, but the experimental integration of magnetized plasma rods within closed systems has not been seen.

\begin{figure}
	\includegraphics[width = 0.7\linewidth]{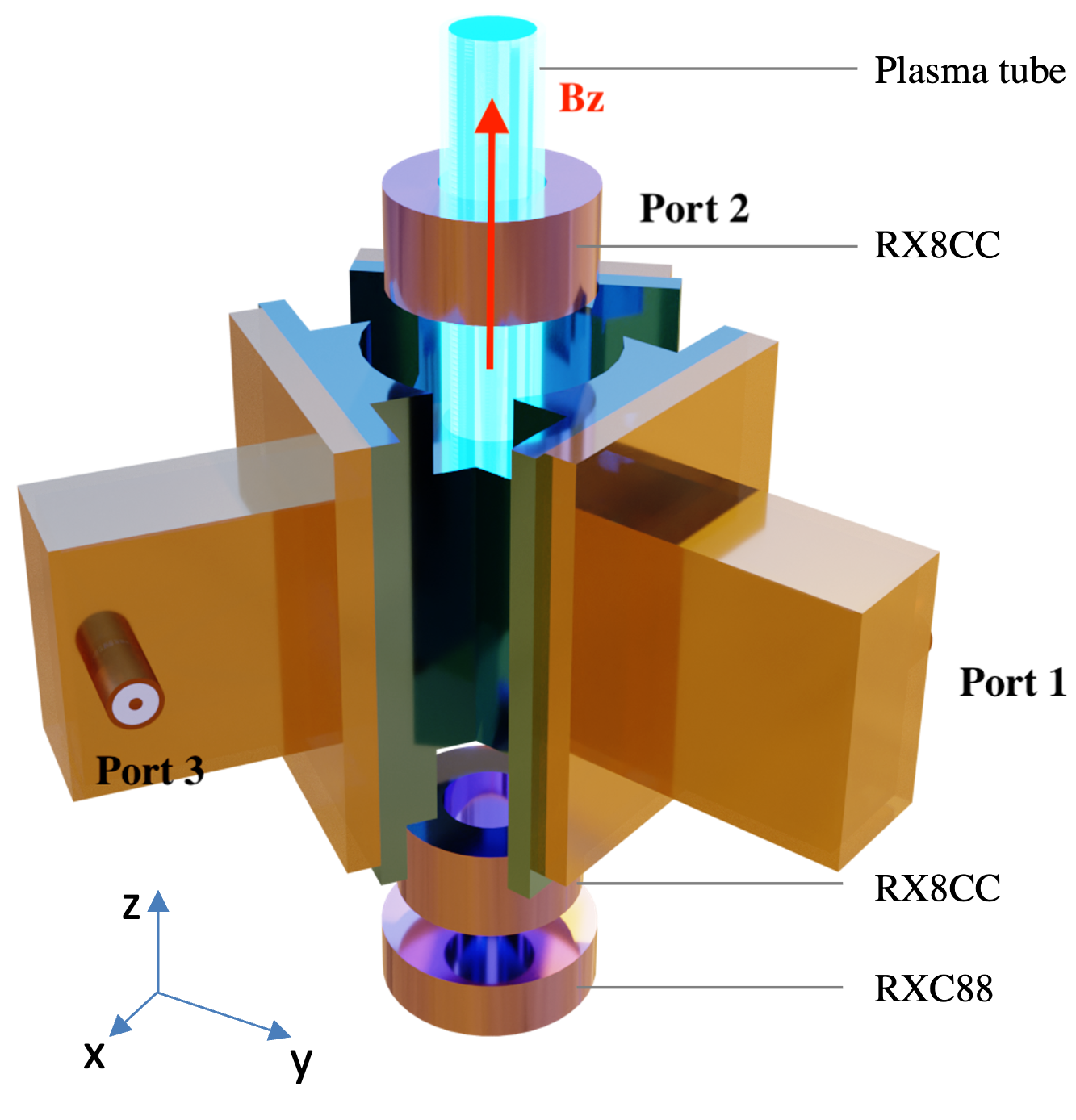}
	\caption{An expanded visualization of the three-port Y-junction circulator. The circulator has three WR187 rectangular-to-coaxial waveguide adaptors (in gold). The neon blue cylinder represents the plasma column. Ring magnets are placed around the tube on the top and bottom of the structure (in purple).}
	\label{fig:setup}
\end{figure}

In this paper, we experimentally demonstrate an E-plane microwave circulator based on a magnetized plasma rod and verify its configurability with plasma parameters and the magnetic field profile (Fig. \ref{fig:setup}). As a proof-of-concept, we evaluate the performance of the device using the same metrics as used for ferrite circulators, though here we are concerned with the verification of the underlying nonreciprocal interaction. 

\section{Wave Propagation in Plasmas}

\begin{figure*}[htpb]
    \includegraphics[width = 0.95\linewidth ,height=0.27\linewidth]{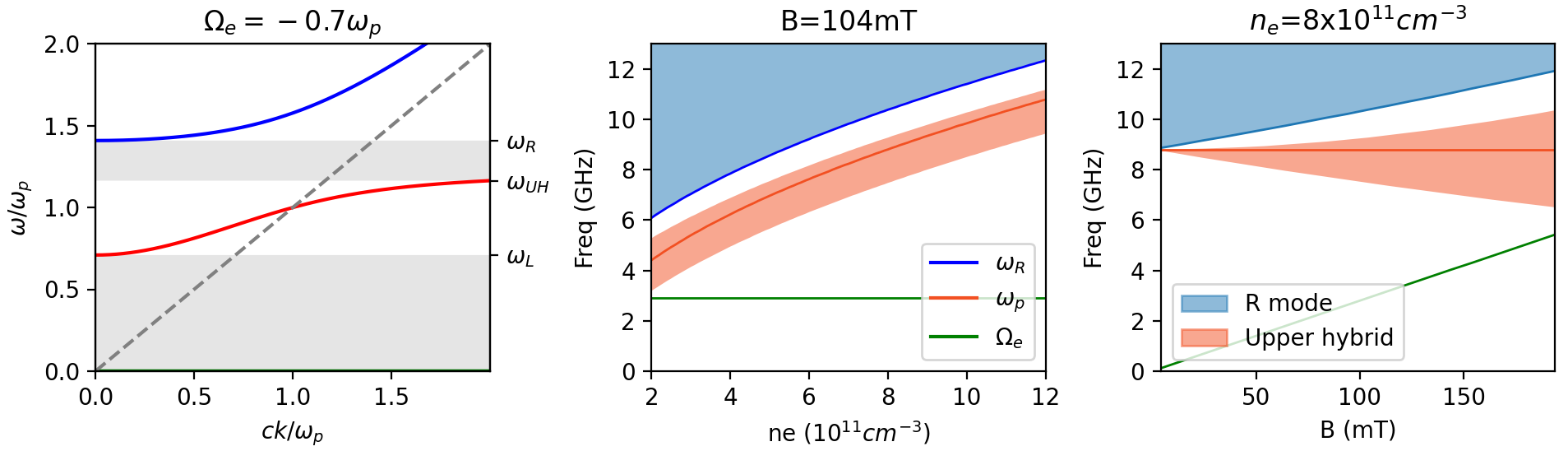}\\

    (a)\hspace{0.33\linewidth}(b)\hspace{0.33\linewidth}(c)
	\caption{(a) The band diagram for X-mode wave propagation in a magnetized collision-less plasma. Solid lines are propagating modes. The dashed grey line is the light line $\omega=kc$. Shaded regions are band gaps. The evolution of the characteristic frequencies with plasma density (b) and magnetic field strength (c) calculated from Eq. \eqref{eq:charw}. The blue region marks frequencies above $\omega_R$ . The red region covers frequencies between $\omega_L$ and $\omega_{UH}$. }
	\label{fig:theory}
\end{figure*}
As a collection of ions and electrons, plasma exhibits varying material properties susceptible to its characteristic parameters and background environments, making it a promising platform for tunable devices. For a simplified model, we assume the collective plasma fluid to be an unbounded, homogeneous, quasi-neutral gas. By making these assumptions, we neglect the pressure gradient at the plasma boundary that causes the charged fluid to flow and assume the ion charge density is equal to the electron charge density, $n_i\approx n_e$. In addition, since ions are much heavier than electrons and respond much slower to electromagnetic field perturbations in the GHz frequency range, we ignore the interaction between ions and waves such that ions only serve as a static scattering background. For a low-frequency wave or a 'hot' plasma with a high collision rate, a rigorous thermodynamics analysis of the ion and electron fluids should be made \cite{boyd_textbook_2007}.

In a non-magnetized cold plasma, under the influence of an incoming electromagnetic (EM) wave, the charged particles move to counteract the field disturbance for wave frequencies lower than the plasma frequency, reflecting most of its energy \cite{boyd_textbook_2007}. However, when a magnetic field is present, electromagnetic waves with specific polarizations can penetrate through the plasma. For this study, we are specifically interested in the X-mode wave, whose electric field is simultaneously perpendicular to the direction of propagation and the applied magnetic field. In this case, the charged particles gain angular momentum under the Lorentz force. The critical feature used in this study is the gyrotropic response, which makes it useful for building a circulator or for time-reversal symmetry breaking in photonic topological insulators \cite{haldane_possible_2008}.  

Considering an X-mode wave propagating inside an unbounded isotropic cold plasma with wavevector $\bm{k}=(k\sin\theta,0,k\cos\theta)$ and angular frequency $\omega$, the wave equation can be expressed by \cite{boyd_textbook_2007, ginzburg_textbook_1970},
\[(\mathbf{k} \cdot \mathbf{E}) \mathbf{k}-k^{2} \mathbf{E}+\left(\frac{\omega}{c}\right)^2\bm{\varepsilon_r} \cdot \mathbf{E}=0,\]
\begin{equation}
\bm{\varepsilon_r}=\left[
\begin{array}{ccc}\varepsilon_m & -i\varepsilon_k & 0 \\ i\varepsilon_k & \varepsilon_m & 0 \\ 0 & 0 & \varepsilon_p\end{array}\right] 
,\quad 
\begin{array}{l}
\varepsilon_m\approx1-\frac{\omega_{p}^{2}(\omega-j\nu_c)}{\omega\left((\omega-j\nu_c)^{2}-\Omega_{e}^{2}\right)},\\
\varepsilon_k\approx\frac{\omega_{p}^{2} \Omega_{e}}{\omega\left((\omega-j\nu_c)^{2}-\Omega_{e}^{2}\right)}\\
\varepsilon_p\approx 1-\frac{\omega_{p}^{2}}{\omega(\omega-j\nu_c)}
\end{array}
\label{eq:xmodedisp}
\end{equation}
where we have defined the coordinate system in such a way that the $z$ axis is along the $\bm{B}$ field and the $y$ axis is normal to the plane containing $\bm{k}$ and $\bm{B}$ with $\theta$ being the angle between them. $c$ is the speed of light in vacuum, $\omega_p$ is the electron plasma frequency, $\Omega_{e}$ is the electron cyclotron frequency, and $\nu_c$ is the electron collision frequency. These frequencies are determined by several factors including the plasma electron density $n_e$, the applied magnetic flux density $B$, the elastic collision cross-section $\sigma$, and the electron velocity $v_{e}$ \cite{ginzburg_textbook_1970}. Specifically, 
\begin{equation}
\omega_p=\sqrt{\frac{e^2n_e}{m_e\varepsilon_0}},\quad \Omega_{e}=-\frac{e B}{m_{e}},\quad \nu_c\propto n_e\langle\sigma v_{e}\rangle
\label{eq:plasmapar}
\end{equation}

Fig. \ref{fig:theory}(a) shows the dispersion relationship of the X-mode wave traveling in a collision-less isotropic plasma. The lower hybrid propagating mode dominated by ions has been neglected in this case. The upper hybrid propagating mode dominated by electrons starts from the left cutoff at $\omega_L$ until it reaches resonance at $\omega_{UH}$. At the resonance, the phase velocity $v_p=\omega/k$ of the wave becomes zero, so that the wave stops propagating due to strong interaction with the charged particles. Above the right cutoff $\omega_R$, the plasma becomes transparent as the electrons cannot respond fast enough to the incoming wave. For a collision-less plasma, the characteristic frequencies can be estimated from the solution of Eq. \eqref{eq:xmodedisp} \cite{boyd_textbook_2007},

\begin{equation}
\begin{array}{rlr}
 \omega_{LH} &\ll \omega_p  &\text{(Lower Hybrid) }\\
  \omega_{UH} &\approx \sqrt{\omega_p^2+\Omega_{e}^2} &\text{(Upper Hybrid) }\\
 \omega_{R,L} &\approx \frac{1}{2}\left(\sqrt{4\omega_p^2+\Omega_{e}^2}  \pm |\Omega_{e}|\right)&\text{(Right/Left Cutoff) }
\end{array}
\label{eq:charw}
\end{equation}

The characteristic frequencies are calculated from Eq. \eqref{eq:charw} for different settings. Fig. \ref{fig:theory}(b) suggests that at constant magnetic field strength, increasing plasma density moves the upper hybrid mode to the higher frequency region with little change in the bandwidth. On the other hand, Fig. \ref{fig:theory}(c) shows that increasing the magnetic field strength under a constant plasma density would broaden the upper hybrid band and the bandgap above.

\section{The Scattering Matrix}

The scattering matrix or S parameters are a simple but effective tool in analyzing microwave transmission through passive linear components. It is widely used to describe the propagation of electromagnetic energy from sources at different ports. For a three-port network, the incident and reflected signal powers are related by the scattering matrix, $\mathbf{S}$,
\[\mathbf{b}=\mathbf{S}\ \mathbf{a}\]
\begin{equation}
\left(\begin{array}{c} 
b_1\\
b_2\\
b_3\\
\end{array}\right)
=\left(\begin{array}{ccc} 
S_{11}&S_{12}&S_{13}\\
S_{21}&S_{22}&S_{23}\\
S_{31}&S_{32}&S_{33}\\
\end{array}\right)
\left(\begin{array}{c} 
a_1\\
a_2\\
a_3\\
\end{array}\right)
\label{eq:S_param}
\end{equation}
where $a_{i}$ is the source power at port i and $b_{j}$ is reflected power at port j for $i,j \in \{1,2,3\}$.

For a network exhibiting Lorentz reciprocity, interchanging the source and the receiving ports should not change the behavior of the network. In this case, the scattering matrix is symmetrical and $S_{ij}=S_{ji}$ for any two ports i and j. On the other hand, an asymmetrical scattering matrix indicates a nonreciprocal system, such as a circulator. The nonreciprocity in a circulator is usually introduced by nonreciprocal materials like magnetized ferrites or magneto-optics materials. It was shown that off-diagonal terms in the material permittivity or permeability tensors can introduce asymmetry in the scattering matrix and thus breaking the local time-reversal symmetry \citep{asadchy_nonrecip_2020}\citep{lax_button_1962}.

An ideal three-port lossless circulator, depending on the direction of circulation, has the scattering matrix in the form of,
\begin{equation}
\mathbf{S}
=\left(\begin{array}{ccc} 
0&0&1\\
1&0&0\\
0&1&0\\
\end{array}\right) \textit{ or } 
\left(\begin{array}{ccc} 
0&1&0\\
0&0&1\\
1&0&0\\
\end{array}\right) 
\label{eq:idealcirc}
\end{equation} 
Considering a single source incident on port one of the circulator described above, the return loss is defined as the fraction of the input power reflected when the other ports are perfectly matched, i.e. $b_1/a_1=S_{11}=0$. The insertion loss is defined as the output power normalized by the input power, i.e. $b_2/a_1=S_{21}=1$. Without a magnetized plasma column, there will be no difference between waves transmitted to port two and port three. Therefore, we defined the effective isolation as the power ratio between the isolated port and the output port and used it to characterize the performance of the plasma circulator in this paper, $\Delta S=|S_{output}| \text{ (dB)}-|S_{isolate}|\text{ (dB)} $. 

\section{Plasma Microwave Circulator}
\label{sec:PlasmCirc}
\subsection{Platform design}
\label{subsec:platform}

To satisfy the boundary conditions for the X-mode wave propagation, we design an E-plane Y-junction structure made of aluminum alloy with standard rectangular waveguide channels. A visualization of the setup is illustrated in Fig. \ref{fig:setup}. The fundamental mode supported by this structure has its electric field vector and the direction of propagation in the $x-y$ plane so $\theta=90^\circ$. The Y-junction is connected to a four-port Vector Network Analyzer (VNA) through three WR187 waveguide-coaxial adaptors. The adaptors support wave propagation in the range of 3.95 - 5.85 GHz with a voltage standing wave ratio less than 1.25. The design is optimized to support the desired frequency range and ensure roughly 68\% of spatial overlap between the propagating mode and the plasma tube. The specifications of the components are summarized in Table \ref{tab:component_geometry}. 

\begin{table}[htp]
    \centering
    \begin{tabular}{p{1.6cm}p{1.2cm} p{5.8cm}}
        \toprule
        Component & Model No. &Specification\\
        \midrule
        Waveguide & WR187  & 1.872 in x 0.872 in, 3.95-5.85 GHz\\
        Plasma &G4T5 & 15.5 mm OD x 13.5 cm Long, 1 mm thick quartz wall\\
        Magnets & RXC88 & 4.45 cm OD x 1.27 cm ID x 1.27 cm thick\\
        &RX8CC  & 3.8 cm OD x 1. 9cm ID x 1.9 cm thick\\
        \bottomrule
    \end{tabular}
    \caption{The geometries of the components used.}
    \label{tab:component_geometry}
\end{table}

\begin{figure*}[ht]
	\includegraphics[width = \linewidth]{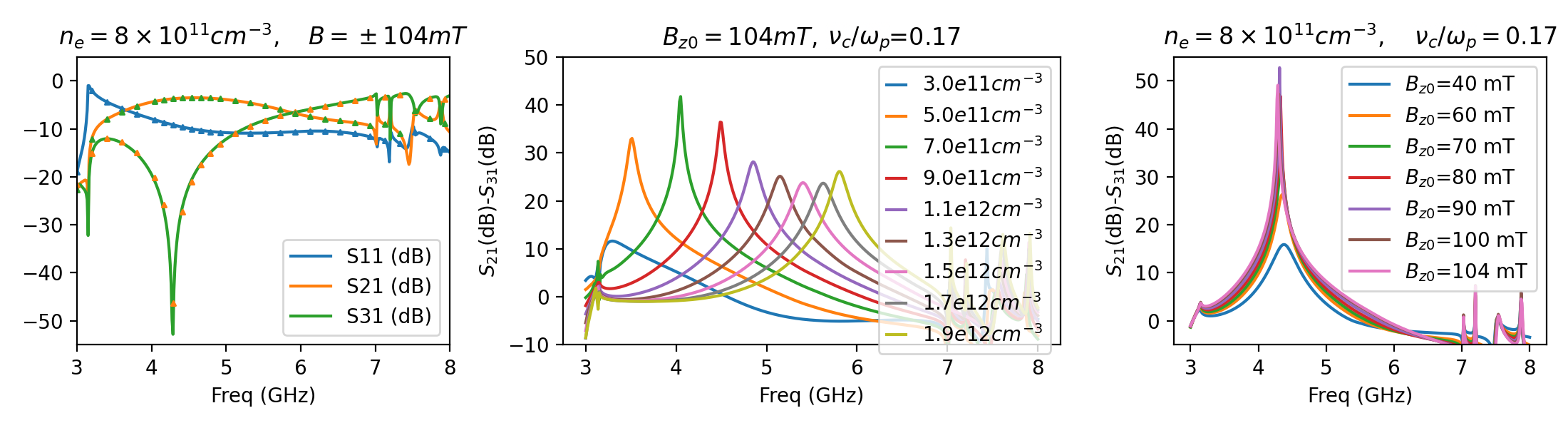}\\
	(a)\hspace{5cm} (b)	\hspace{5cm} (c)
	\caption{(a) The simulated S parameters when $n_e=8\times 10^{11}$ cm$^{-3}$, $\nu_c/\omega_p=0.17$, B=104 mT (solid lines) and B=-104 mT (triangle marker). The change of the effective isolation peak with (b) the plasma density (c) the magnetic field strength.}
	\label{fig:sim_neB}
\end{figure*}

For the plasma source, a Ushio G4T5 germicidal UV lamp filled with low-pressure argon gas and mercury drops is used. When a high voltage is applied, the tungsten filaments of the electrodes heat up rapidly and produce electrical discharge through ionized argon gas. The mercury atoms are vaporized and excited when they collide with the accelerated electrons. UV light at 253.7 nm is emitted when the atoms are de-excited \cite{lama_analytical_1982}. The lamp is powered by a variable transformer that is adjustable from 0 to 148 V. An AC ballast operating at 33.1 kHz \cite{luc_2022} is connected in series to regulate the power drawn by the negative resistance of the lamp. In laboratory settings, a glow discharge can usually sustain a plasma with electron density of $10^{10} \sim 10^{12}$ cm$^{-3}$ , corresponding to plasma frequencies of $f_p=\omega_p/2\pi=1 \sim 9$ GHz \cite{Conrads_2000}.

The design strategy for the magnetic field is to maximize the field at the center by using a combination of neodymium ring magnets, minimizing the distance between them, and using thicker magnets with larger cross-sectional areas. In this setup, the magnets are separated by a minimum distance of 5.2 cm, so long as the UV lamp does not obstruct the path and passes through the ring. Therefore, we place RXC88 and RX8CC ring magnets on the bottom and another RXC88 magnet on the top for the lamp to pass through. Using a 3-axis hall effect sensor, the axial magnetic field at the center is measured to be 104 mT $\pm$ 1 mT. The radial component is less than 11 mT in magnitude and varies no more than 5 mT across the diameter. 

\subsection{Simulation}
\label{subsec:sim}

A simplified model of the circulator is simulated with ANSYS HFSS driven mode solver using the Finite Element Method (FEM). The circulator is assigned an aluminum boundary with wave port excitation at each of the three ends. By doing so, we de-embedded the coaxial-waveguide transition in the simulation. The plasma tube is simulated by defining a homogeneous anisotropic material with a frequency-dependent permittivity tensor calculated from Eq. \eqref{eq:xmodedisp}. In addition, a quartz envelope of 1 mm thickness is placed around the plasma. The background material is air. Parametric sweeps on the collision frequency, plasma density, and magnetic field strength are performed to study wave propagation through the circulator.

\begin{figure}[hpb]
	\centering
	\includegraphics[width = \linewidth]{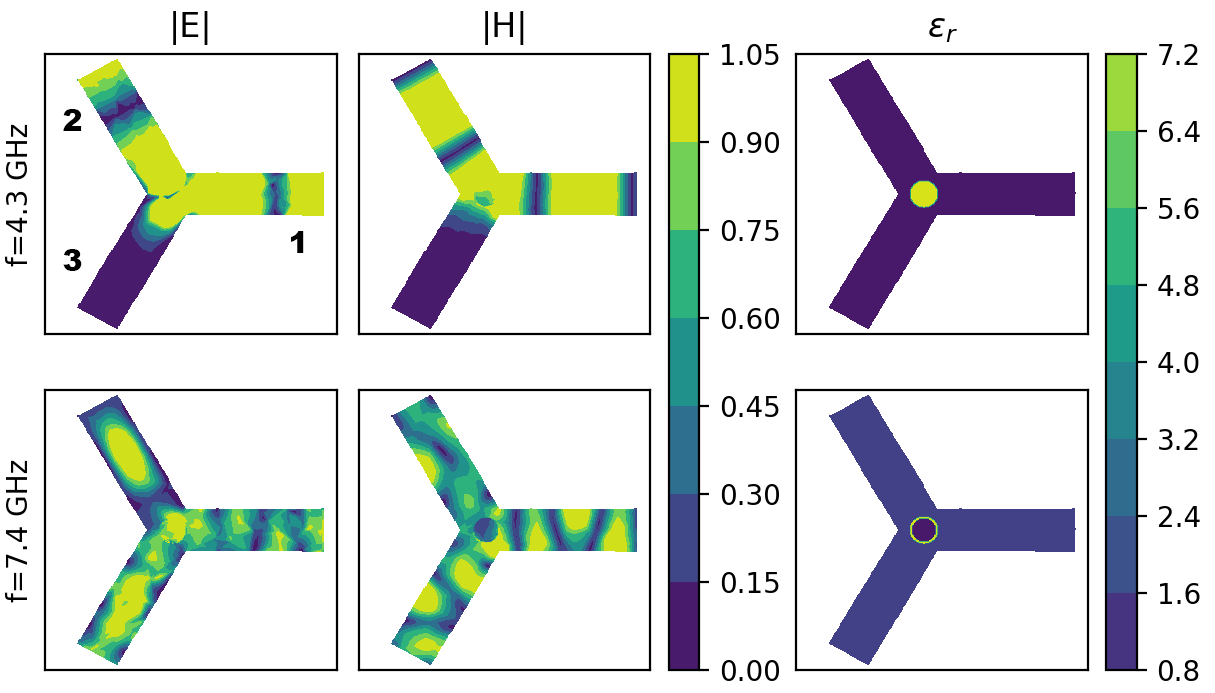}	\\
	(a) \hspace{2cm} (b) \hspace{2cm} (c)
	\caption{The normalized magnitude of the E field (a), H field (b) and the approximate relative permittivity profiles (c) at 4.3 GHz and 7.4 GHz for the solution in Fig. \ref{fig:sim_neB}(a) when B=104 mT. Port numbers were marked as in the top left subplot.}
	\label{fig:sim_expBnB}
\end{figure}

An example of the driven solver solution when $B=\pm$ 104 mT, $n_e=8\times 10^{11}$ cm$^{-3}$, and $\nu_c/\omega_p=0.17$ is shown in Fig. \ref{fig:sim_neB}(a). As expected, inverting the direction of the magnetic field switches the output and the isolation ports. For the positive magnetic bias case, the normalized isolation power ($|S_{31}|$) can be as low as -52 dB at 4.3 GHz. The insertion loss ($|S_{21}|$) is -3.5 dB and the 10 dB effective isolation bandwidth is around 1 GHz. Plasma absorption and impedance mismatch are the two main factors to account for the insertion loss. At maximum isolation, around 46\% of the power is absorbed by the plasma. The normalized field magnitude and relative permittivity are plotted in Fig. \ref{fig:sim_expBnB} to confirm the circulator behaviour at 4.3 GHz and equal power splitting at 7.4 GHz.

\begin{figure*}[htb]
	\includegraphics[width = 0.95\linewidth]{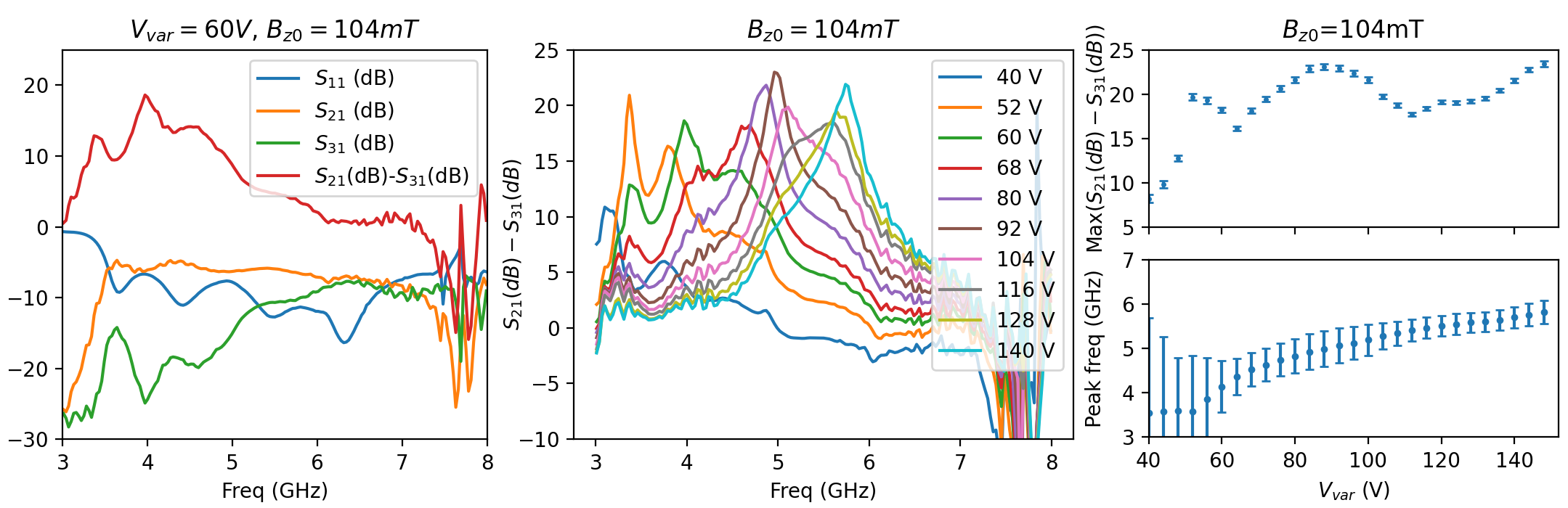}\\
	(a)\hspace{5cm} (b) \hspace{5cm} (c)
	\caption{The experimental results. (a) The S parameters when $B_{z0}=104$ mT and the variable transformer is set to $V_{var}=60$ V. (b)  The effective isolation vs frequency. The peak shifts to the right with higher excitation voltage. (c) Peak values, extracted from the experimental data. The peak effective isolation (top) and the frequency at the peak (bottom) vs the voltage of the transformer.}
	\label{fig:exp_100mT_ne}
\end{figure*}
By changing the ratio between the collision frequency and the plasma frequency, we found that the collision frequency varies the amplitude of the effective isolation peak (see Appendix Fig. \ref{fig:sim_vc}). In contrast, it has little effect on the return loss or the peak isolation frequency. The optimum $\nu_c/\omega_p$ for maximum isolation is correlated to the applied magnetic field strength. Since it is difficult to accurately measure and reconfigure the collision frequency with non-invasive methods, we use the optimum value $\nu_c/\omega_p=0.17$ for the simulation so that the simulation results can serve as a reference performance for a perfectly matched circulator.

Assuming the same $\nu_c/\omega_p$ ratio, Fig. \ref{fig:sim_neB}(b) suggests that increasing plasma density will shift the effective isolation peak to higher frequencies. The 10 dB bandwidth is roughly 1.0 $\pm$ 0.1 GHz for plasma density ranging from $5\times 10^{11}$ to $11.5\times 10^{11}$ cm$^{-3}$. Outside of this range, part of the isolation peak resides outside of the operating band of the waveguide. Furthermore, Fig. \ref{fig:sim_neB}(c) shows that with the same collision and plasma frequencies, the overall effective isolation peak increases with the magnetic field strength, whereas the peak isolation frequency only shifts by a negligible amount.

\subsection{Experimental results}
The VNA is calibrated to the end of the coaxial cable, and all measurements are taken with a time averaging of 20 frames. When the lamp is turned off, the measured return loss ($|S_{11}|$) of the radio-frequency (RF) signal fluctuates in the waveguide operation band with a mean value of 7.6 dB. As discussed previously, the high reflection could be due to the impedance mismatch at the center of the Y-junction. In addition, the coaxial-rectangular mode transition varies the power coupled into the circulator for different frequencies. Despite the high reflection, the difference in the normalized power transmission to the two output ports, $S_{21}$ and $S_{31}$, is always below 0.3 dB within the frequency range of interest.

Assuming the plasma is generated uniformly inside the lamp of radius $R_a$, the plasma's electron density can be related to the current and voltage of the UV lamp through current density $J$,
\begin{equation}
\begin{split}
J=e \langle n_e \bm{v_d}\rangle=I/(\pi R_a^2),\quad n_e \propto I
\label{eq:current}
\end{split}
\end{equation}
where $v_d$, the electron drift velocity, depends on the electron mobility $\mu_e$ and the electric field strength $E$ that is proportional to the electric potential. The $I_{rms}$ passing through the lamp increases with the voltage output from the transformer as measured in Appendix Fig. \ref{fig:UV_IV}. Therefore, to study the effect of the plasma density on the circulator performance, we measure the S parameters when $B_{z0}=104$ mT while changing the output voltage of the variable transformer from 40 to 148 volts. The lamp is always ignited at high voltage first to warm up the electrodes for a stable excitation. 

Fig. \ref{fig:exp_100mT_ne}(a) shows the S parameters when the transformer is sending power at 60 V. Looking along the $z$-axis, the device demonstrates a right-hand clockwise (CW) circulation with the effective isolation peaked at 18.6 dB around 4.1 GHz. At least 10 dB of effective isolation is observed for 40\% bandwidth from 3.3 to 5.0 GHz. The averaged power transmission is summarized in Table \ref{tab:my_label} for the bandwidth inside the waveguide operation band. The averaged signal loss from the system $L=1-\sum_{i=1}^{3}|S_{i1}|$ is 57\% due to the absorption inside the plasma. 

\begin{table}[htpb]
    \centering
    \begin{tabular}{cccc}
    \toprule
        \text{power} & $|S_{11}|$ (dB)  &$|S_{21}|$ (dB) &$|S_{31}|$ (dB)  \\
    \midrule
    \text{off}&-7.6&-4.1&-4.2\\
    \text{on}&-8.7&-5.7&-15.7\\
    \bottomrule
    \end{tabular}
    \caption{The mean value of the S parameters when the lamp is turned off and when it is turned on at 60 V with $B=B_{z0}=$ 104 mT in the range of 4 - 5 GHz.}
    \label{tab:my_label}
\end{table}

\begin{figure}[htpb]
	\includegraphics[width = 0.75\linewidth]{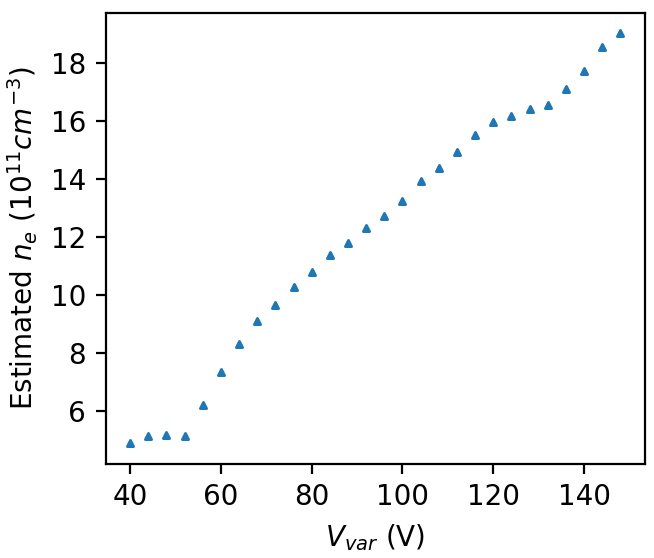}
	\caption{The estimated plasma charge density at different transformer voltage ouputs for $B_{z0}$=104 mT.}
	\label{fig:104mT_estne}
\end{figure}

Fig. \ref{fig:exp_100mT_ne}(b) summarizes how the effective isolation changes with the voltage output from the transformer. The effective isolation peak has a 10 dB bandwidth of around 1 GHz and increasing the voltage moves the peak to higher frequencies without altering its bandwidth significantly. This resembles the simulation results in Fig. \ref{fig:sim_neB}(b). We can therefore associate the plasma density of the lamp to the voltage from the transformer by matching the peak isolation frequencies obtained from the experiment and the simulation. For voltage below 52 V, a large portion of the isolation peak sits outside the band of the waveguide adaptors causing fluctuations near the band edge. To avoid finding the local maxima, the center frequencies of the overall peak envelope are extracted by fitting the data with a Lorentz distribution. The results are plotted in Fig. \ref{fig:exp_100mT_ne}(c) with the parameter variance as the errors. By matching the peak isolation frequencies, the plasma density of the UV lamp is estimated to vary between $5\times 10^{11}$ to $19\times 10^{11}$ cm$^{-3}$ for transformer voltages ranging from 40 to 148 V (Fig. \ref{fig:104mT_estne}). The order of magnitude of the estimated plasma density is in good agreement with other estimates \cite{luc_2022}. Since the lamp used in this study has higher r.m.s current and similar radius, our estimation is reasonably higher than the values in the reference. 

\begin{figure*}[ht]
	\includegraphics[width = 17.2cm]{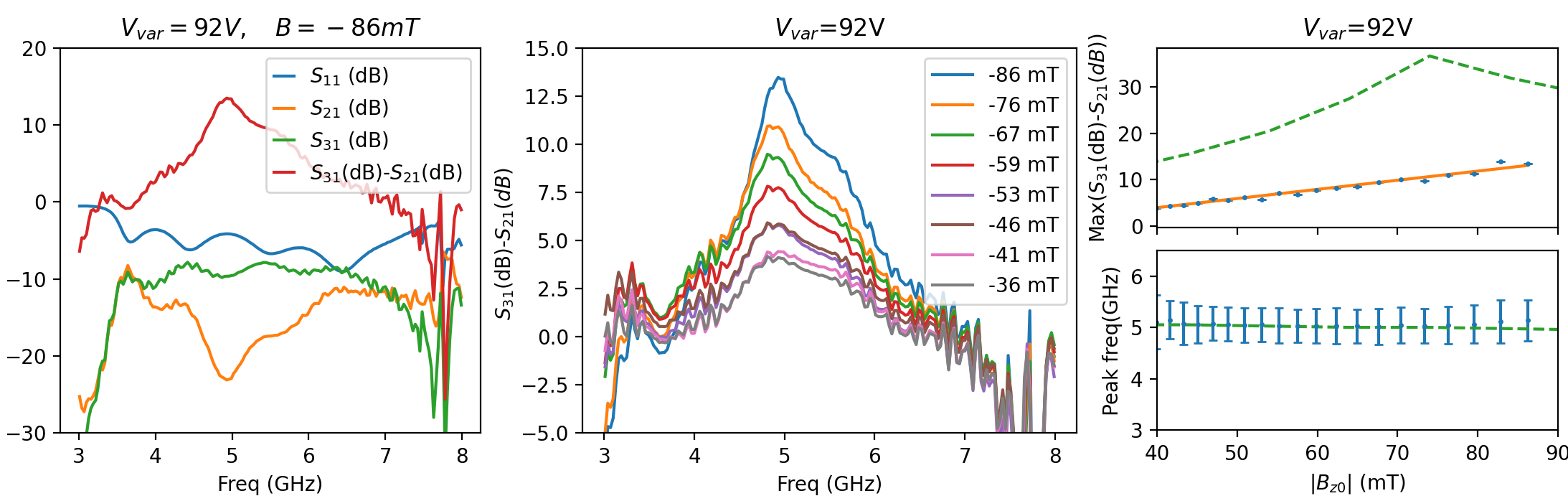}\\
	(a)\hspace{5cm} (b) \hspace{5cm} (c)
	\caption{The experimental results. (a) The S parameters when the variable transformer is set at 92 V. (b) The effective isolation vs frequency at different $B_{z0}$. (c) Peak isolation magnitude and frequency variation. Blue error bars are data extracted from (b). The green dashed line are results from the simulations when $n_e=11.5\times 10^{11}$ cm$^{-3}$, $\nu_c/\omega_p=0.17$. The orange solid line is the linear regression between the maximum effective isolation and the center magnetic field strength with $R^2\approx 0.99$.}
	\label{fig:exp_B}
\end{figure*}
To study the effect of $B$ on the isolation peak, we vary the center magnetic field strength by introducing plastic spacers in increments of 0.5 mm thickness. Since it is difficult to experimentally reconfigure the field strength without shifting the field center, we remove the RXC88 magnet to keep $\bm{B}$ symmetric about the center. The center field strength decreases exponentially from 86 mT to 34 mT by adding spacers with a combined thickness of 0 to 23 mm (see Appendix Fig. \ref{fig:Bvsd}). The transformer output is kept fixed at a single value between 64 V to 120 V.

The S parameters are plotted in Fig. \ref{fig:exp_B}(a) when the variable transformer is set to 92 V and the center magnetic field is set to -86 mT by reversing the direction of the magnets. As expected, the wave propagation is rerouted to port 3 instead of port 2 when the field direction is flipped. In Fig. \ref{fig:exp_B}(b-c), we observe that empirically, the amplitude of the isolation peak increases proportionally to the magnetic field strength. However, the position of the peak center does not shift significantly. Fig. \ref{fig:exp_B}(c) also shows that the peak isolation frequency extracted from the experiments is in good agreement with that from the simulation when $n_e=11.5\times 10^{11} cm^{-3}$ for all field strengths.

Multiple factors can explain the discrepancy in the magnitude between the simulations and the experiments. First of all, the actual collision frequency could be different from the resonant collision frequency used in the simulation. In addition, the mode mismatch during the coaxial-rectangular transition can degrade the isolation because of high reflection. Furthermore, the simulation assumes an isotropic magnetic field and a homogeneous plasma rod which would not be the case for actual experiments. The spatial inhomogeneity in the field profile could spread the peak power over a range of frequencies due to dispersion. 

Finally, from both simulations and experiments, we observe band shifting with the plasma density and amplitude increasing with the magnetic field strength. These phenomena demonstrate interesting similarities to the evolution of the upper hybrid mode in Fig. \ref{fig:theory}(b). However, if we plot the peak isolation frequencies at different plasma densities from the simulation onto Fig. \ref{fig:theory}(b), they lie about halfway through the bandgap region below the upper hybrid mode. The key differences between these models are the presence of waveguides and collisions. Limited by the scope of this paper, these phenomena will be further detailed in future works.

\section{Discussion and Conclusion}
We have demonstrated nonreciprocal transmission of electromagnetic waves through magnetized plasma through an E-plane Y-junction circulator using a commercially available germicidal bulb and permanent ring magnets. Nonreciprocal transmission is observed in the range of 4 to 6 GHz both in simulations and experiments with the circulation direction chosen by the direction of the applied magnetic field. The nonreciprocal transmission is an essential proof of using magnetized plasma as an experimental platform for plasma integrated photonic topological insulators with versatile tunable options. Specifically, the original demonstrations \cite{wang_observation_2009} for topologically non-trivial devices employed gyromagnetic materials, as most solid state gyroelectric materials have a Voigt parameter less than 1 at room temperature under sub-tesla field. Within magnetized plasmas, this value can be much higher under the same conditions and can be tuned by several parameters (magnetic field strength, electron density, etc.). 

The design of the circulator involves optimizing the geometry for laboratory-achievable plasma densities and magnetic field strengths. For an effective circulator, the return loss needs to be lower than the insertion loss so that enough power is transmitted to the preferred port. In the low voltage regime, plasma absorption dominates the transmission loss. In the high voltage regime, more particles are ionized increasing the plasma and the collision frequencies, leading to more reflection and power loss through heat. As a result, the excitation voltage for the plasma needs to be carefully balanced for high isolation and low return loss. For this specific design, the circulator can reach up to 18.6 dB effective peak isolation at 4 GHz with a 10 dB bandwidth of 1.6 GHz when the variable transformer is set to 60 V and the center magnetic field strength is 104 mT. The performance could be further improved by enlarging the waveguide structures to support lower plasma frequencies and increasing the magnetic field strength to increase the total isolation. Furthermore, it is possible to reduce the return loss at the designated frequency by adding a matching structure at the center, such as a circular cavity or step structures. 

From both the simulations and the experiments, we found that with our design, the operating frequency band of the circulator is primarily determined by the plasma density. At the same time, the strength of the magnetization mostly controls the magnitude of the peak isolation. By matching the peak isolation frequency from the simulations and the experiments, we estimate the full range plasma density of the UV lamp to be $5\times 10^{11} \sim 19 \times 10^{11}$ cm$^{-3}$ when it is powered by a 40 to 148 V AC transformer. The FEM simulation shows good reliability as a design tool for optimizing the circulator's performance. 

The simple plasma circulator design, although without impedance matching, already shows comparable isolation and bandwidth to commercially available microwave ferrite circulators. It indicates a good nonreciprocity ($>10$ dB) for GHz bandwidth. However, the insertion loss is more that of a low-loss ferrite circulators (less than -1 dB). This might raise problems for using plasma inside a circulator but the primary aim of the project focuses more on exploring a new platform for nonreciprocity than designing a new circulator device. Ferrite materials are known to be bulky and brittle. Their response is limited by their saturation magnetization, usually in the range of sub-tesla field strength. On the other hand, plasma can lead us in various experimental directions. Different discharge mechanisms can generate charge densities from $10^{8}$ cm$^{-3}$ to $10^{14}$ cm$^{-3}$ for interaction with a wide range of frequencies. The spatial charge density profile can be varied by factors such as the pressure, shape of the electrodes, or the confinement magnetic field. The fast response of charged particles to their environment makes plasma particularly useful for designing actively tunable devices. In addition, this study can be beneficial for waveguiding in a plasma concentrated environments. The nonreciprocal microwave transmissions in cases of a highly magnetized plasma or a high-power input signal (inducing nonlinear effects) are also interesting topics that are yet to be explored in great detail.

\begin{acknowledgements}

The authors acknowledge the support from the U.S. Air Force Office of Scientific Research through the Multidisciplinary Research Program of the University Research Initiative (MURI Grant NO. FA9550-21-1-0244). The authors thank Dr. H. Bernety, L. Houriez, J. Rodriguez, Dr. B. Wang, and Professor M. Cappelli from Stanford University for generously providing us with the plasma tubes and the ballasts and for our fruitful discussions on the topic of characterizing plasma parameters. 
\end{acknowledgements}

\appendix
\section{Extended Data}

\begin{figure}[htp]
	\includegraphics[width =\linewidth]{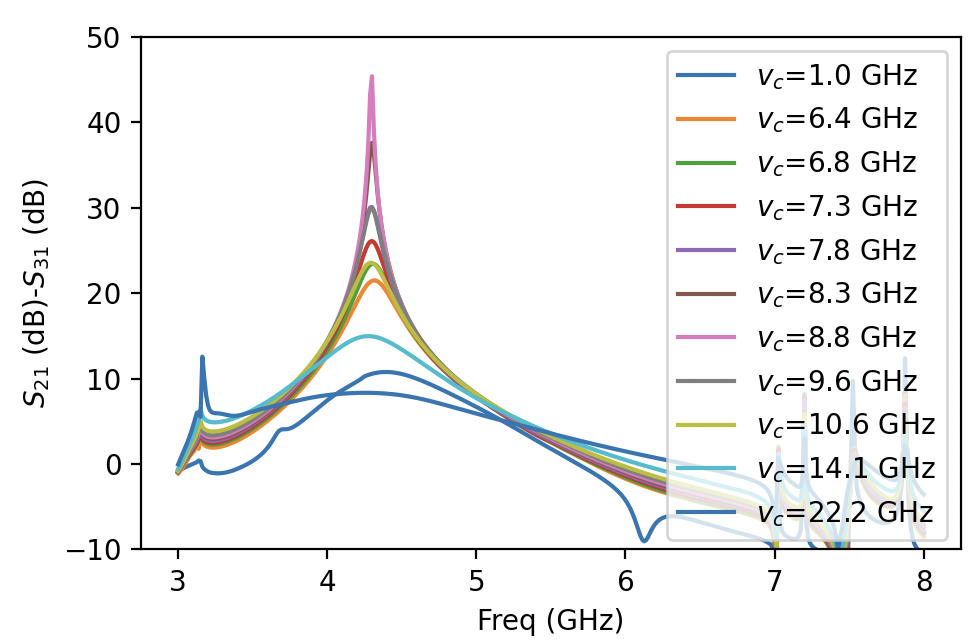}	\\
	\caption{Simulation result: The effective isolation peak at different collision frequencies when $n_e=8\times 10^{11}$ cm$^{-3}$, $B=104$ mT.}
	\label{fig:sim_vc}
\end{figure}

A parametric sweep of the simulated circulator performance when changing the collision frequency for a magnetized plasma with $n_e=8\times 10^{11}$ cm$^{-3}$ and $B=104$ mT is shown in Fig. \ref{fig:sim_vc}. An optimum collision frequency $\nu_c=8.8$ GHz gives the maximum effective isolation at 4.3 GHz. However, it is challenging to manipulate the collision frequency in experiments since it is closely dependent on factors such as plasma density, the lamp temperature, and the strength of the magnetic bias. As a result, a quantitative study on the collision frequency is not included in this work.

\begin{figure}[htp]
    \centering
    \includegraphics[width=\linewidth]{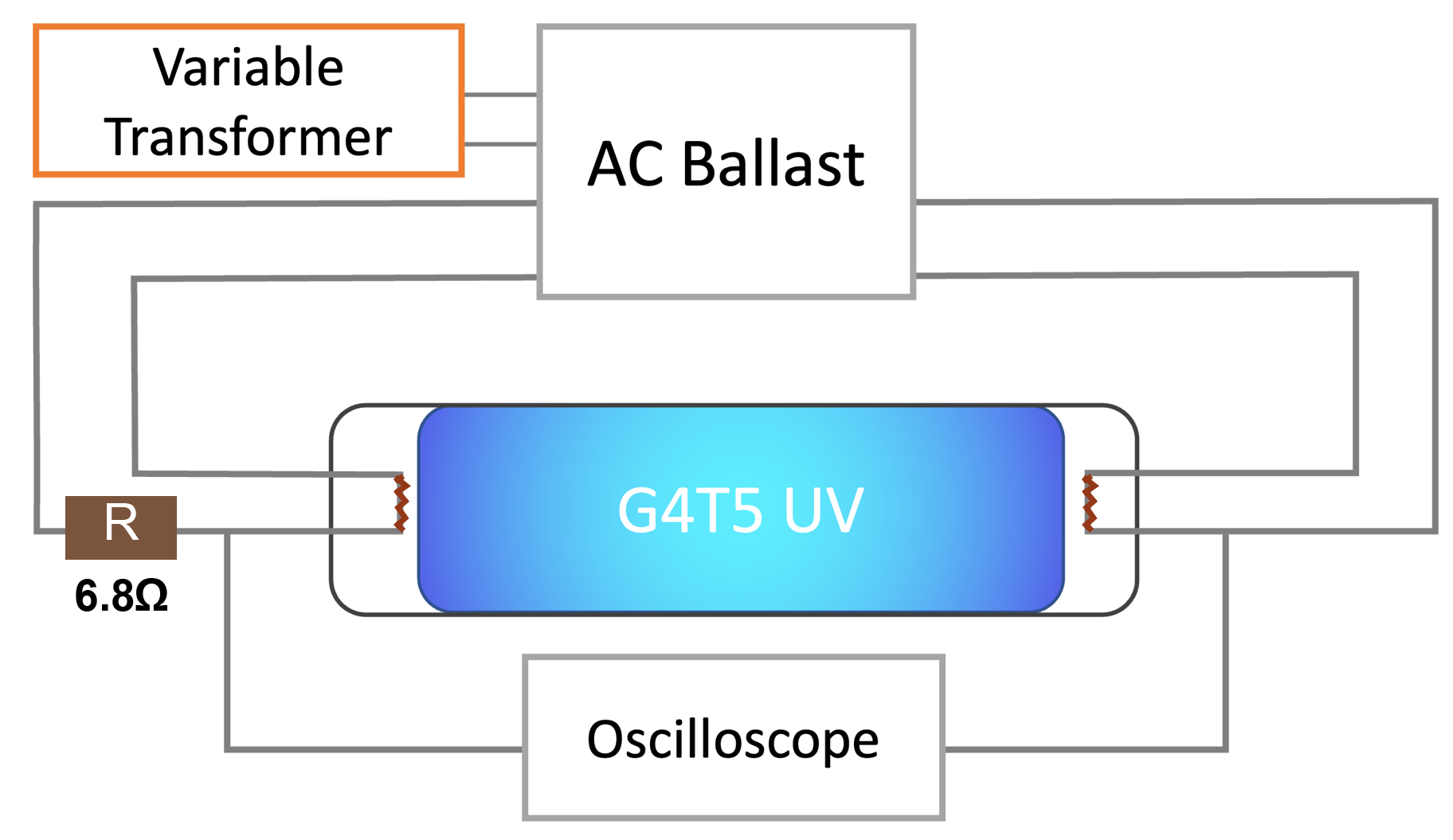}
    \caption{A schematic circuit diagram for the UV lamp}
    \label{fig:lamp_circuit}
\end{figure}

As shown in Fig. \ref{fig:lamp_circuit}, the UV lamp is connected to a variable transformer through a 33.1 kHz AC ballast. The differential voltage across the lamp is measured with two passive voltage probes using an oscilloscope. The current is measured by the voltage across a shunt resistor of 6.8 $\Omega$ connected in series with the lamp. The r.m.s. parameters are calculated by $U_{rms}=\sqrt{\frac{1}{N}\sum_{i=1}^{N}|U(t_i)|^2}$ with $N$ being the number of samples in time, $t_i$ being the time indices and $U$ being the parameters to be measured.

\begin{figure}[htp]
	\includegraphics[width =\linewidth]{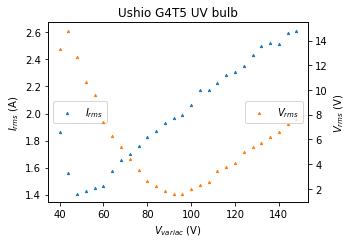}	\\
	\caption{The r.m.s current (blue) and the r.m.s voltage (orange) of the germicidal UV lamp v.s. the output voltage from the transformer. The lamp is measured with Keysight MSO7014B oscilloscope.}
	\label{fig:UV_VI}
\end{figure}

\begin{figure}[htpb]
	\centering
	\includegraphics[width = \linewidth]{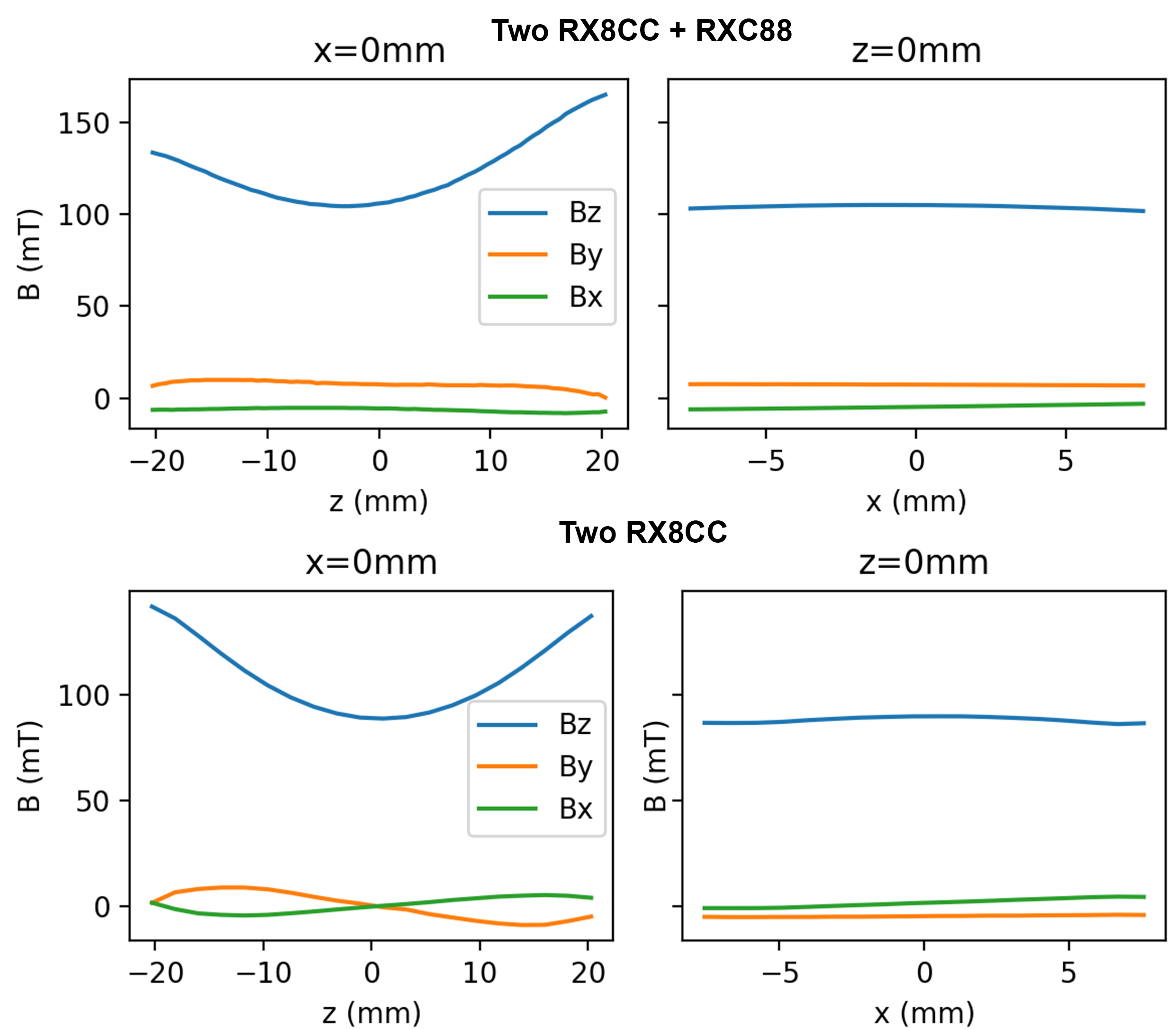}\\
	\caption{The vector components of the magnetic field strength measured along the $z$ axis (left panel) and along the $x$ axis (right panel).}
	\label{fig:UV_IV}
\end{figure}

\begin{figure}[htpb]
    \centering
	\includegraphics[width = 0.9\linewidth]{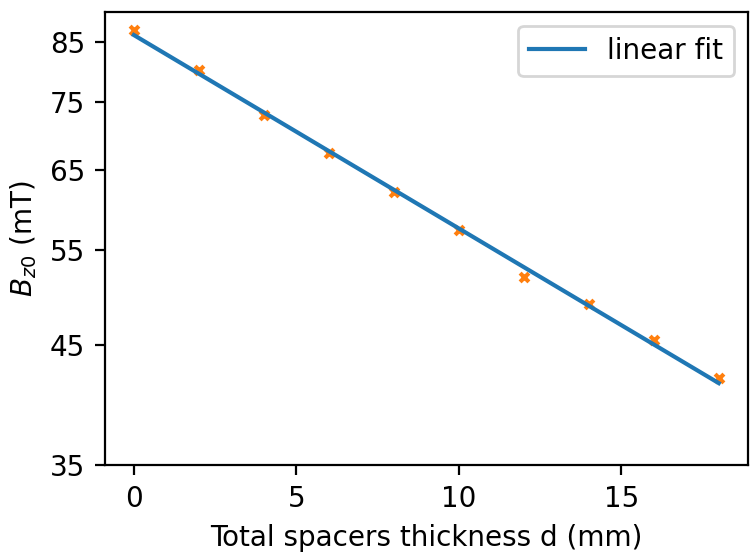}
    \caption{The magnetic field strength at the origin decreases exponentially with the total spacer thickness introduced. The $y$-axis is $B_{z0}$ plotted in logarithm scale. $\ln(B_{z0})=-0.04d+4.46$}
    \label{fig:Bvsd}
\end{figure}

\bibliography{PlasmaCirc}

\end{document}